# Entropic rigidity of a crumpling network in a randomly folded thin sheet

Alexander S. Balankin and Orlando Susarrey Huerta

*Fractal Mechanics group, National Polytechnic Institute, Av. Politécnico Nacional, México D.F., México 07738*

We have studied experimentally and theoretically the response of randomly folded hyperelastic and elasto-plastic sheets on the uniaxial compression loading and the statistical properties of crumpling networks. The results of these studies reveal that the mechanical behavior of randomly folded sheets in the one-dimensional stress state is governed by the shape dependence of crumpling network entropy. Following to the original ideas by Edwards for granular materials, we derive an explicit force-compression relationship which precisely fit the experimental data for randomly folded matter. Experimental data also indicate that the entropic rigidity modulus scale as the power of the mass density of folded ball with the universal scaling exponent.

# I. INTRODUCTION

Random folding of thin materials is of noteworthy importance to many branches of science and industry. Examples range from virus capsids and polymerized membranes to folded engineering materials and geological formations [1,2,3,4,5]. They usually consist of thin sheets or rods constrained to undergo large deformations. Accordingly, the folding phenomena are associated with a rich class of crumpling phenomena [3], which belong to a wider class of interfacial deformation phenomena [5]. Because of their biological and technological importance, the properties of randomly folded thin materials are now the subject of increasingly growing attention [5,6,7,8,9,10,11,12,13,14,15,16,17,18,19,20,21, 22,23,24,25,26,27,28,29,30,31,32,33,34,35,36,37,38,39,40,41,42,43,44,45].

Formally, folding of self-avoiding matter is a continuum of isometric embeddings of the $d$-dimensional manifold in the n-dimensional space [46]. A rich variety of self-generated configurations in randomly folded materials is governed by their dimensionality, the constitutive nature of deformations, and the nature of the forces causing the deformation [5,32,37]. While randomly folded materials are examples of ill-defined systems, because the folding procedures appear quite haphazard, the experiments with randomly folded thin sheets are rather well reproducible [37,42,43,47], because of the topology and self-avoiding interactions are two most important physical factors when dealing with folding of thin materials [32,37,47]. In this way it was found that despite the complicated appearance of folded configurations, the folding phenomenon is in itself very robust, because almost any thin material crumples in such a way that the most part of folding energy (>90%) is concentrated in the network of narrow crumpling creases (ridges) that meet in the point-like vertices [5,6,48]. The properties of ridges have been studied thoroughly. Scaling laws governing the energy and size of the ridge have been obtained analytically [5,49] and tested numerically [9,32,48,50] and experimentally [36,37,43]. Furthermore, it was shown that the balance of bending and stretching energy in the crumpling creases determines the scaling properties of the folded state as the function of the confinement force, sheet dimensions, and mechanical properties of thin material [5,6,9,32].

Specifically, numerical simulations of random folding with a coarse-grained model of triangulated self-avoiding surfaces with bending and stretching elasticity [32] suggest that the characteristic size of folded configuration $R$ scales with the hydrostatic confinement force $P$ as

$$\frac{R}{h} \propto \left(\frac{L}{h}\right)^{2/D} \left(\frac{P}{Eh}\right)^{-\delta_3}, \qquad (1)$$

where $h$ and $L$ are the thickness and edge size of sheet ($h \ll L$), $E$ is the two-dimensional Young's modulus of sheet, $\delta_3$ is the folding force scaling exponent, and $D$ is the fractal dimension of the set of elastic sheets with different edge sizes folded by the same confinement force $P = const$. Namely, according to the scaling behavior (1), a set of randomly folded thin sheets of the same thickness but different edge size $L$ is expected to obey a fractal law

$$L^2 \propto R^D , \qquad (2)$$

when all sheets are folded by the same hydrostatic force $P = const$. The fractal scaling behavior (2) was observed in numerous experiments with randomly folded papers [35,37, 43,47,51,52,53,54,55,56], metal foils [18,35,42,57,58], and cream layers [38].

The numerical simulations suggest that for randomly folded self-avoiding two-dimensional elastic sheets the scaling exponent $\delta_3 = 1/4$ and the fractal dimension $D = 2.3$ are universal [32]. Experimentally, it was found that in the case of predominantly plastic deformations of folded sheets, such as aluminum foils and cream layers, the fractal dimension $D = 2.3 \pm 0.1$ is independent on the sheet thickness and the folding force and consistent with the universal value found in numerical simulations [38,42]. However, in experiments with different kinds of elasto-plastic paper the fractal dimension $D$ is found to be material dependent [37,51-56]. The later was attributed to the strain relaxation in randomly folded elasto-plastic sheets after the folding force is withdrawn [37].

More recently, it was found that the internal structure of folding configurations also possesses scaling invariance within a wide range of length scales [43]. The fractal dimension of folding configurations is found to be universal $D_l = 2.64 \pm 0.05$, *i.e.* independent on sheet thickness and material properties [43], and close to the fractal dimension $D = 8/3$ expected for a randomly folded phantom sheet with finite bending rigidity [32]. This finding implies that the self-avoidance does affect the scaling properties of the internal structure of randomly folded thin matter (see [43]). The scaling behavior with $D < D_l$ was termed as an intrinsically anomalous self-similarity [43].

Statistical properties of crumpling networks formed in randomly folded materials were also studied theoretically [3,9,15,30], in experiments [25,37,44], and by numerical simulations [30,32]. It was found that crumpling networks are disassortative [44] and exhibit statistical self-similarity [37]. However, there is no consensus about the statistical distribution of crumpling crease sizes. Theoretical considerations [3,30], numerical simulations [30,32], and experimental studies [37] suggest that the crease length distribution obeys a log-normal distribution at relative low confinement of folded sheet and gamma distribution at higher confinement ratios, whereas more recently, the authors [44] have reported and power-law functions to give good fits for this distribution. Furthermore, it was suggested that crumpling network governs the mechanical behavior of randomly folded materials [9,32,48], which exhibit anomalously low compressibility under hydrostatic pressure [17,32,36,37].

Generally, the mechanical response of any network is determined by the volume and shape dependence of its free energy [59]. However, numerical simulations [32] and experiments [17,36] suggest that the mechanical behavior of randomly folded sheets in a thee-dimensional stress state ($k=3$) is dominated by the volume dependence of crumpling network enthalpy, $U$, leading to the power-law force-compression relation:

$$P = F_3 = R^{-1}\left(\frac{\partial U}{\partial \lambda}\right) \propto Y_3 R \lambda^{-1/\delta_k} \tag{3}$$

where $\lambda = r/R$ is the compression ratio, $R$ and $r(F_3)$ are the characteristic size of randomly folded sheet before and after the deformation, respectively; and

$$Y_k = \left(\frac{\partial F_k}{\partial r}\right)_{\lambda=1} \tag{4}$$

is the mechanical stiffness of folded sheet under k-axial compression in a three-dimensional stress state. Furthermore, numerical simulations suggest that the folding force scaling exponents $\delta_k \leq 1/2$ take only the universal values determined by the corresponding universality classes [32]. Specifically, numerical simulations of self-avoiding sheet folding with a coarse-grained model of triangulated surfaces with bending and stretching elasticity suggest the following relationship for the folding force scaling exponent [32]

$$\delta_k = 1/(k+1), \tag{5}$$

while in experiments with randomly folded aluminum foils it was found $\delta_3 = 0.21 \pm 0.02$ [36]. Further, it was found that under the uniaxial and radial compressions randomly folded thin sheets exhibit Poisson's expansion obeying a power-law behavior with the universal Poisson's index $\nu = 0.17 \pm 0.01$ [60]. However, the mechanical behavior of randomly folded thin materials under non-hydrostatic forces remains poorly understood.

Though in the three-dimensional stress states the entropic contribution to the force-deformation behavior of crumpling network is negligible, in a one-dimensional stress state, the shape dependence of crumpling network entropy may play a major role, as it is for the stretching of folded proteins [61]. However, while the entropic elasticity of flexible networks is by now well understood [59,61,62,63,64,65], the entropic contribution to the crumpling network rigidity was not studied yet. Accordingly, to gain insight into the mechanical behavior of crumpling network in the one-dimensional stress-state, in this work we studied the mechanical behavior of randomly folded thin sheets under uniaxial compression.

**II. EXPERIMENTS**

In this work we studied the effect of crumpling network on the mechanical behavior of randomly folded thin sheet. For this purpose we tested hand folded sheets of elasto-plastic papers of different thickness and hyperelastic latex rubber. In both cases, it is expected that the mechanical response of folded material is governed by the crumpling network. However, if rubber sheet crumpling is completely reversible, the stress concentration in the crumpling ridges leads to plastic deformations of paper. As a result, the large deformations of randomly folded paper are essentially irreversible (see Fig. 1). This limits the applicability of equilibrium thermodynamic for describing the mechanical behavior of randomly folded paper. On the other hand, uncontrolled unfolding of rubber sheets makes difficult the study of randomly folded rubber under axial compression (see Fig. 2). Furthermore, there is no way to study the statistical properties of crumpling network in randomly folded rubber sheets. Accordingly, to study the effect of crumpling network on the mechanical behavior of randomly folded sheets we used different kinds of paper. Experiments with rubber sheets were performed to confirm that the strain-stress relaxation in elasto-plastic paper (see refs. [37]) does not affect the essential nature of the force-compression behavior of folded matter under the uniaxial compression with loading rates used in this work.

## 2.1. Materials tested

To study the statistical properties of crumpling networks and their effect on the mechanical behavior of randomly folded thin sheets, we used square sheets of three commercial papers of different thickness, $h = 0.024 \pm 0.004$, $0.039 \pm 0.003$, and $0.068 \pm 0.005$ mm, early used in refs. [37,43]. The edge size of square paper sheets $L$ was varied from $L_0 = 4$ to 66 cm with the relation $L = qL_0$ for scaling factor $q = 1, 2, 2.5, 4, 5, 7.5, 8.75, 9, 10, 15$ and $16.5$. The paper sheets were folded in hands into approximately spherical balls. At least 30 balls with different confinement ratios $K = L/R$ were folded from sheets of each size of each paper. Once the folding force is withdrawn, the ball diameter increases with time during approximately 6–9 days, due to the strain relaxation (see, for details, ref. [37]). So, all experiments reported below were performed at least ten days after a sheet was folded.

The mean diameter of each ball $R$ was determined from measurements along 15 directions taken at random. Early, it was found that the folded configurations of randomly folded papers are characterized by the universal local fractal dimension $D_l = 2.64 \pm 0.05$ [43], whereas the fractal dimension $D$ of the set of balls folded from sheets with different size $L$ is the thickness dependent (see ref. [37]). Early we reported that the sets of balls folded from papers of thickness $h = 0.024\pm0.004$, $0.039\pm0.003$, and $0.068\pm0.005$ mm obey the fractal behavior (2) with the fractal dimensions $D = 2.13\pm0.05$, $2.30\pm0.05$, and $2.54\pm0.06$, respectively [37]. In this work we obtain the same results.

In contrast to paper, the deformations of latex rubber are completely reversible. We used two square sheets of latex rubber of thickness 0.1 mm with edge sizes of 150 and 250 mm. Rubber sheet was folded in hands and fixed between the clams to avoid unfolding (see Fig. 2 a) just before the mechanical test. In this way, five uniaxial compression tests were performed with the sheets of each size.

**2.2. Mechanical tests**

At least 10 balls folded from sheets of each size of each paper were tested under axial compression with the compression rates of 0.1 mm/sec. using a universal test machine (see Fig. 1). Additionally, the sets of paper balls folded from sheets of sizes 30x30 and 60x60 cm$^2$ were testes at the compression rates of 1 mm/sec. Furthermore, ten experiments were performed with balls folded from latex rubber (see Fig. 2).

Figure 3 shows a typical force – compression behaviour of randomly folded paper ball under uniaxial compression. While the deformations of folded paper are essentially irreversible, we found that the loading part of the force-compression curve $F_1(\lambda = H/R)$ does not depend on the compression rate, at least in the range used in this work. At the same time, we noted that the force – compression behaviour does not obey the power-law scaling (3) (see insert in Fig. 3). Moreover, we found that in all cases the loading part of experimental force-compression curve $F_1(\lambda)$ may be precisely fitted (see Figs. 3 and 4) by the following simple relationship

$$F_1 = -Y_1\left(\frac{1-c}{\lambda-c}-1\right) \qquad (6)$$

for $\lambda > c$, where the fitting parameter ($c \ll 1$) and the stiffness (4) of folded ball under uniaxial compression ($Y_1 \ll Y_3$) are independent on the compression rate. Specifically, we found that the fitting parameter scales with the folded ball size as

$$c \propto R \qquad (7)$$

(see Fig. 5 a), while the rigidity modulus

$$E = \frac{Y_1}{R^2} = R^{-2}\left(\frac{\partial F_1}{\partial r}\right)_{\lambda=1} \qquad (8)$$

scales with the mass density of folded ball,

$$\rho = \frac{\rho_0 6hL^2}{\pi R^3}, \qquad (9)$$

as

$$\frac{E}{E_0} = \left(\frac{\rho}{\rho_0}\right)^\phi, \qquad (10)$$

where $\rho_0 = 900 \pm 50$ kg/m$^3$ is the mass density of papers (see [37]) and $E_0$ is the material dependent constant (see Fig. 5 b). At the same tame, we found that the data for different papers presented in Fig. 5 b are best fitted with the power law function (10) with the scaling exponent $\phi = 2.1 \pm 0.1$ [66].

Further, we found that the equation (6) also provides the best fit of the stress-compression behaviour of randomly folded rubber sheets under axial loading (see Fig. 6). Unfortunately, we were not able to perform the systematic studies of fitting parameters with respect to folding conditions, because of the problems with uncontrolled unfolding of rubber sheets. Nonetheless, the main conclusion from Fig. 6 is that randomly folded hyperelastic sheets obey the same force-compression relation (6) under uniaxial loading as it found for randomly folded elasto-plastic paper. So, we can assume that the nature of mechanical response of randomly folded thin sheets under uniaxial loading is independent on the nature of bending deformations of sheet.

Above we already noted that the force-compression behaviour (6) drastically differs from the power-law force-compression behaviour (3) associated with the volume dependence of network enthalpy. Furthermore, we noted that the mechanical rigidity of folded matter under uniaxial compression is much less than under hydrostatic compression, i.e., $Y_1 \ll Y_3$, and the elastic modulus scaling exponent $\phi = 2.1 \pm 0.1$ [67] is close to the universal value $\phi = 2.2$ expected for the scaling exponent of the entropic rigidity modulus (see ref. [61]). So, taking into account the low bending rigidity of thin sheets, one can expect that the mechanical response of randomly folded material on the uniaxial compression is primarily determined by shape dependence of crumpling network entropy, rather than the volume dependence of sheet energy.

**2.3. Statistical properties of folded sheets and crumpling networks**

In paper, crumpling creases leave permanent marks, and so the crumpling network can be easily visualized after unfolding [37,44]. Accordingly, in this work, ten balls of each size of papers with the thickness 0.039±0.003, and 0.068±0.005 [67] were carefully unfolded and scanned to study the statistical properties of crumpling network. To better visualization of crumpling network, each crumpling crease was marked with pencil during the unfolding process. The scanned images (see Fig. 7 a) were used to reconstruct crumpling networks formed by straight ridges which meet in the point-like vertices (see Fig. 7 b). In total 220 crumpling networks were analyzed. The statistical distributions of measured parameters were determined with the help of @RISK4.5 software [68].

Specifically, in this work we studied the statistical distribution of ridge lengths ($l$). Previous works have reported log-normal [3,30,32,37], gamma [30,37], and power-law [44] functions to give good fits for this distribution. In this work, making the use of the chi-square, Anderson-Darling, and Kolmogorov-Smirnov tests for goodness of fit, performed with the help of the @RISK software [68], we found that the crumpling ridge length distribution in sheets folded with low confinement ratios $K < 4$ is best fitted by the log-normal distribution (see Fig. 8 a), whereas in sheets folded with high confinement ratios $K > 6$ the best fit of ridge length distribution is given by the Gamma-distribution (see Fig. 8 b),

$$P(l) = \frac{m^m}{\Gamma(m)} \left(\frac{l}{l_m}\right)^{m-1} \exp\left(-m\frac{l}{l_m}\right), \quad (11)$$

where

$$l_m \propto R \quad (12)$$

is the mean ridge length (see ref. [36]), $\Gamma(.)$ is the gamma function, and $m$ is the shape parameter, which is proportional to the number of crumpling ridges $N_r$ (see ref. [30, 69]). The last determines the number of layers (see Fig. 9 a) in randomly folded ball $n \propto f(N_r)$, where $f(x)$ is an increasing function of $x$ (see refs. [30,36]). Under the assumption that each layer is incompressible (because the compressibility of paper is much small that of the folded ball) the number of layers can be also estimated as $n = H_{\min}/h$, where $H_{\min}$ is the minimal thickness of ball under axial compression $F \to \infty$. Experimentally, we found that under axial compression of $F = 40$ kN, the number of layers behaves as

$$n \propto R^2 \quad (13)$$

as it is shown in Fig. 9 b.

## III. DISCUSSION

Randomly folded sheets show very general reproducible mechanical behavior characterized by a few control parameters. Early, it was shown that the energy balance between elastic bending and stretching energies in a crumpling ridge is responsible for the rigidity of cylindrical plates and spherical shells [5] and, accordingly, is central to a fundamental understanding of deformation such as in folding of sheets and membranes. Specifically, the crumpling network determines the mechanical behaviour of randomly folded thin sheets under external forces [32].

The thermodynamics of networks evolving at equilibrium is well described by statistical mechanics [61-63]. However, the "frozen configurations" of crumpling network in randomly folded matter does not evolve in absence of some external driving [32]. In nature, there are many systems in such "frozen states". Examples range from supercooled liquids quenched at zero temperature in states, called inherent states [70,71], to granular materials in which grains are "frozen" because the thermal kinetic energy is negligible compared to the gravitational energy and so, the external bath temperature can be considered equal to zero [72]. By analogy with granular materials and supercooled liquids, we can treat the mechanically stable "frozen" folding configurations as inherent states. So, we can follow to the original ideas by Edwards for granular materials [73,74,75] and attempt to develop a statistical mechanics approach for the inherent states of crumpling networks along the line of refs. [76,77].

As much as systems of standard statistical mechanics, each macroscopic state of crumpling network corresponds to a huge number of microstates. So, the first step is to individuate the states distribution, namely what is the probability to find the crumpling network in a given inherent state. We can define the configuration space as the set of all configurations or states of the crumpling network permitted by the folding constraints, with paths in the space corresponding to motions (foldings) of the sheet. Further, we can expect that under stationary conditions the crumpling network is "randomized" enough, and therefore, following essentially Edwards original ideas, we make the assumption that such a distribution is given by a maximization of the entropy under the condition that the average energy is fixed. Specifically, we can consider a statistical ensemble of equivalent folded sheets all prepared in the same way. So, we indicate with $\{U_i\}$ the energies of the accessible inherent microscopic states of each crumpling network and with $n_i$ the number of networks with energies equal to $U_i$. The average energy per network is thus $U = \sum_i p_i U_i$, where $p_i = n_i / N$ is the probability to have a network in the inherent state $i$. Accordingly, the configurational entropy is defined as $S = -\sum_i p_i \ln p_i$ [76,77].

The dynamics from one folding state to another can be induced by the external force $F_k$. We assume that the kinetic energy driven in the folded sheet is rapidly dissipated in crumpling ridges and so, the sheet is almost instantaneously "frozen" in one of the inherent folding states. We can treat these states as quasi-stationary, because of the macroscopic properties change very slowly (see [17,37]). Hence, the stationary distribution is given by the maximal entropy under the folding constraint which fixes the average energy. This requirement leads to the Gibbs distribution function $p_i = \exp(-\beta U_i)/Z$, where the partition function $Z = \sum_i \exp(-\beta U_i)$ is a normalization factor and $\beta$ is a Lagrange multiplier determined by the constraint on the energy (see [76,77]). Accordingly, as in standard statistical mechanics, in the thermodynamic limit the entropy is defined as the logarithm of the number of microscopic inherent states $\Omega(U)$ corresponding to the given macroscopic energy $U$ [77]. Namely,

$$S = \ln Z - \beta U = \ln \Omega(U) \qquad (14)$$

where

$$\beta^{-1} = X = \left(\frac{\partial U}{\partial S}\right)_{F_d} \qquad (15)$$

is the "configurational temperature" of crumpling network (see refs. [74-77]). This "temperature" characterizes the equilibrium distribution among the inherent states and depends only on the average energy of the inherent states and not on the particular dynamics used [78].

To determine the entropic response of crumpling network on the uniaxial compression, here we analyzed the shape dependence of crumpling network entropy. Taking into account that the bending rigidity of a thin sheet is much less than its stretching rigidity and the bulk rigidity of sheet material [32], from Eqs. (11) - (14) follows that under the uniaxial compression the change of the network entropy depends on the compression ratio as

$$S \propto (\lambda - c) - \ln(\lambda - c), \tag{16}$$

where

$$c = \frac{nh}{R} \tag{17}$$

is the ratio of the "incompressible" layers to the ball diameter. Hence, the entropic contribution to the force-compression behaviour

$$F_1 = -\left(\frac{X}{R}\right)\left(\frac{\partial S}{\partial \lambda}\right)_X \tag{18}$$

obeys the relationship (6), where the ball stiffness (4) is $Y_1 \propto cX$ and the parameter $c$ is defined by the relationship (17). Taking into account the ball size dependence of $n$ (see Eq. (13)), from Eq. (17) follows the experimentally observed scaling behaviour (7).

## IV. CONCLUSIONS

The main conclusion to be drawn from our studies is that, in the one-dimensional stress state the response of crumpling network to the uniaxial loading is predominantly of the entropic nature. Accordingly, the loading part of force-compression curve of randomly folded sheets display very general reproducible mechanical behaviour (6) characterized by a few control parameters. However, in contrast to the entropic elasticity of molecular networks [61-63], the (slow) stress relaxation due to the plastic deformations in folding creases leads to the irreversibility of force-compression behaviour of folded paper balls (see Figs. 1, 3).

Randomly folded matter is just one example of a broad category of materials which can be found in "frozen states". The stable configurations of crumpling network are the minima or saddle points of the potential energy or, more generally, all folding states which are mechanically stable. Hence, one may expect that entropic rigidity of crumpling network plays an important role in diverse folding processes. So, our findings provide a novel insight into the crumpling phenomena, ranging from the folding of polymerized membranes to the earth's crust buckling.

## ACKNOWLEDGMENTS

This work has been supported by CONACyT of the Mexican Government under Project No. 44722 and the TELMEX company.

**Figure captions**

**Figure 1**. Set up of axial compression test of randomly folded paper: (a-c) loading and (d-e) unloading.

**Figure 2**. Set up of axial compression test of randomly folded latex rubber sheet: (a-c) loading and (d-e) unloading.

**Figure 3**. The force ($F$) – compression ($\lambda = H/R$) curve of the paper ball ($R$=400 mm) under uniaxial compression with the constant displacement rate $\dot{u} = 0.1$ mm/sec. Circles – experimental data, curve – fitting with the equation (6). Upper inserts show the set-up of the experiment (ball folded from the paper of thickness 0.039 mm); lower insert shows the force-compression curve in the log-log coordinates.

**Figure 4**. The loading parts of the force-compression curves in the coordinates $-F$ versus the dimensionless parameter $\Lambda = (1-c)/(\lambda - c)$ for: a) balls of different diameters folded from sheets of paper with the thickness 0.039 mm; b) balls folded from sheets of size 300x300 mm$^2$ of the papers of different thickness. Symbols – experimental data, straight lines – fitting with the equation (6).

**Figure 5**. a) Fitting parameter $c$ (dimensionless) as a function of $R$ for papers of different thickness. Symbols – experimental data, straight lines indicates the scaling behaviour (7). b) Relative modulus of rigidity $E/E_0$ as a function of relative mass

density $\rho/\rho_0$ of balls folded from papers of different thickness. Symbols – experimental data, straight line straight lines indicates a scaling given by the equation (10) with $\phi = 2.1$. Open and filled symbols are corresponded to the experiments with the compression rates of 0.1 and 1.0 mm/sec., respectively.

**Figure 6**. The loading parts of the force-compression curves in the coordinates $-F$ versus the dimensionless parameter $\Lambda = (1-c)/(\lambda -c)$ for randomly folded latex ball under axial compression. Circles – experimental data, straight line data fitting with Eq. (6).

**Figure 7.** a) Scanned image of unfolded sheet of paper and b) the graph of corresponding crumpling network.

**Figure 8.** Probabilistic distributions of ridge length of crumpling networks in sheets of thickness 0.068 mm folded with different confinement ratios: a) $K = 3.8$ (bins – experimental data, solid line – data fitting with the log-normal distribution) and b) $K = 6.1$ (bins – experimental data, dashed line – data fitting with the log-normal distribution, solid line – the best fit with gamma distribution).

**Figure 9.** a) Picture of a cut through a crumpled ball of paper of thickness 0.068 mm and b) log-log plot of the number of layers $n$ versus the initial diameter of randomly folded paper ball ($R/h$) for papers of different thickness: $h = 0.024\pm0.004$ mm (triangles), $0.039\pm0.003$ mm (rhombs), and $0.068\pm0.005$ mm (circles); the slopes of straight lines is 2.

[66] Unfortunately, relative density of balls, which we were able to test in this work, varies over only one decade ( $0.047 \leq \rho/\rho_0 \leq 0.52$ ), and so, strictly speaking a more correct statement is that the experimental data are consistent with the power law scaling (10) with the universal scaling exponent.

[78] Configurational temperature $X$ enters as a parameter in the equilibrium distribution allowing to substitute, as in usual statistical mechanics, time averages, which are theoretically difficult to evaluate, with ensemble averages. In granular materials the dynamics from one stable microstate to another is induced by sequences of "taps", in which the energy is pumped into the system in pulses [71]. Due to inelastic collisions the

kinetic energy is totally dissipated after each tap, and the system is again frozen in one of its inherent states [71,72]. In the case of crumpling network under uniaxial compression the role of taps play the buckling of crumpling creases.

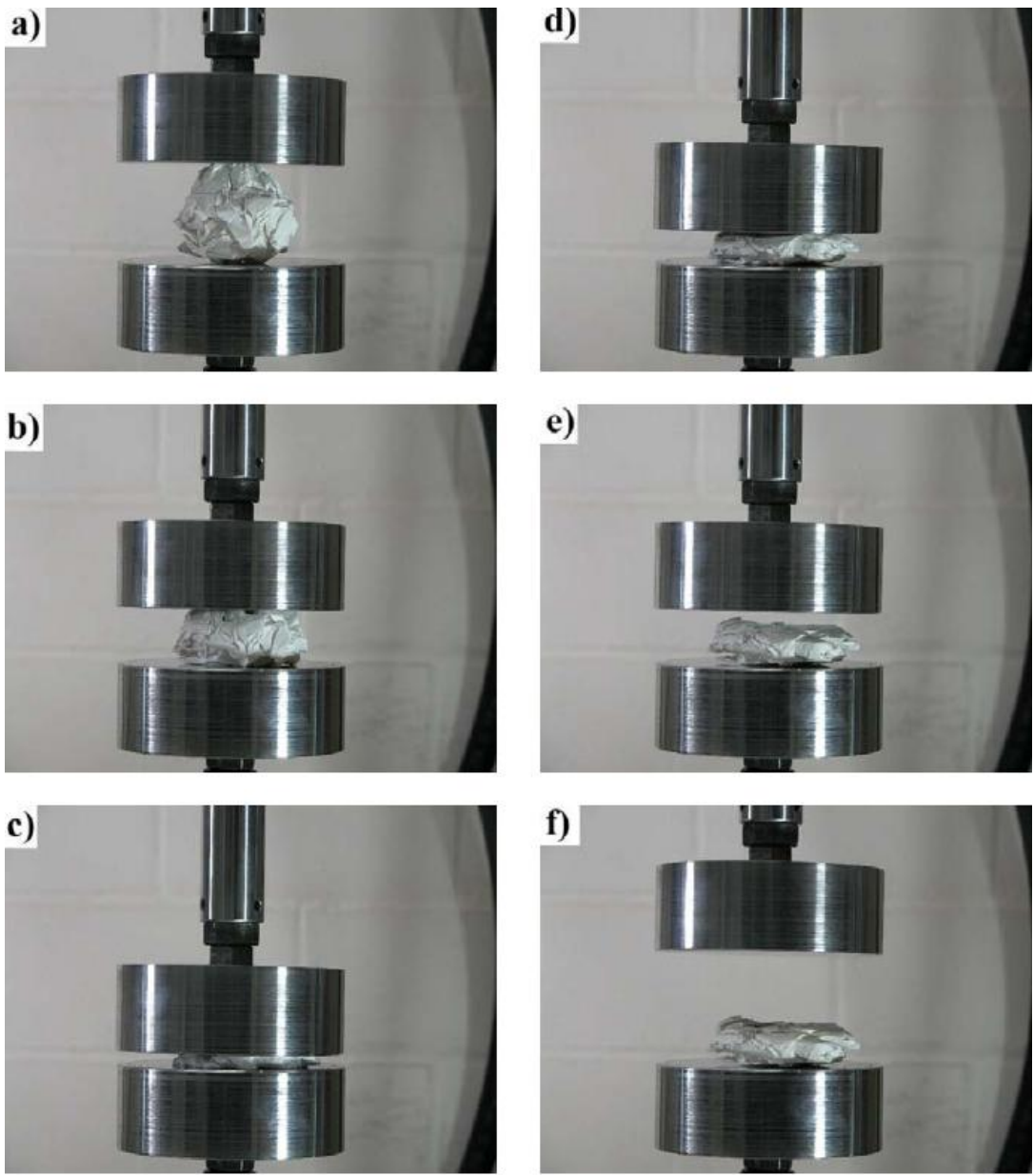

Figure 1.

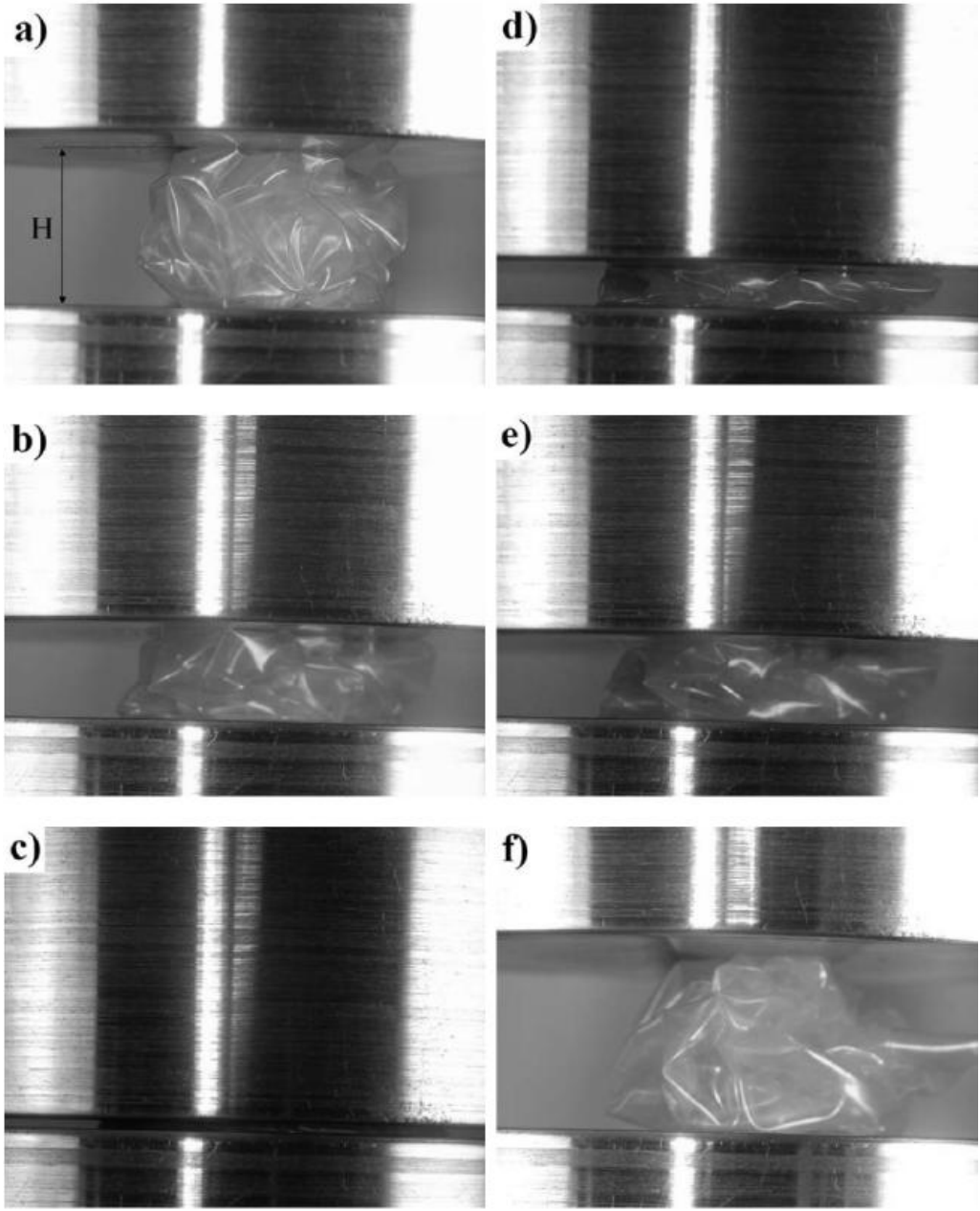

Figure 2.

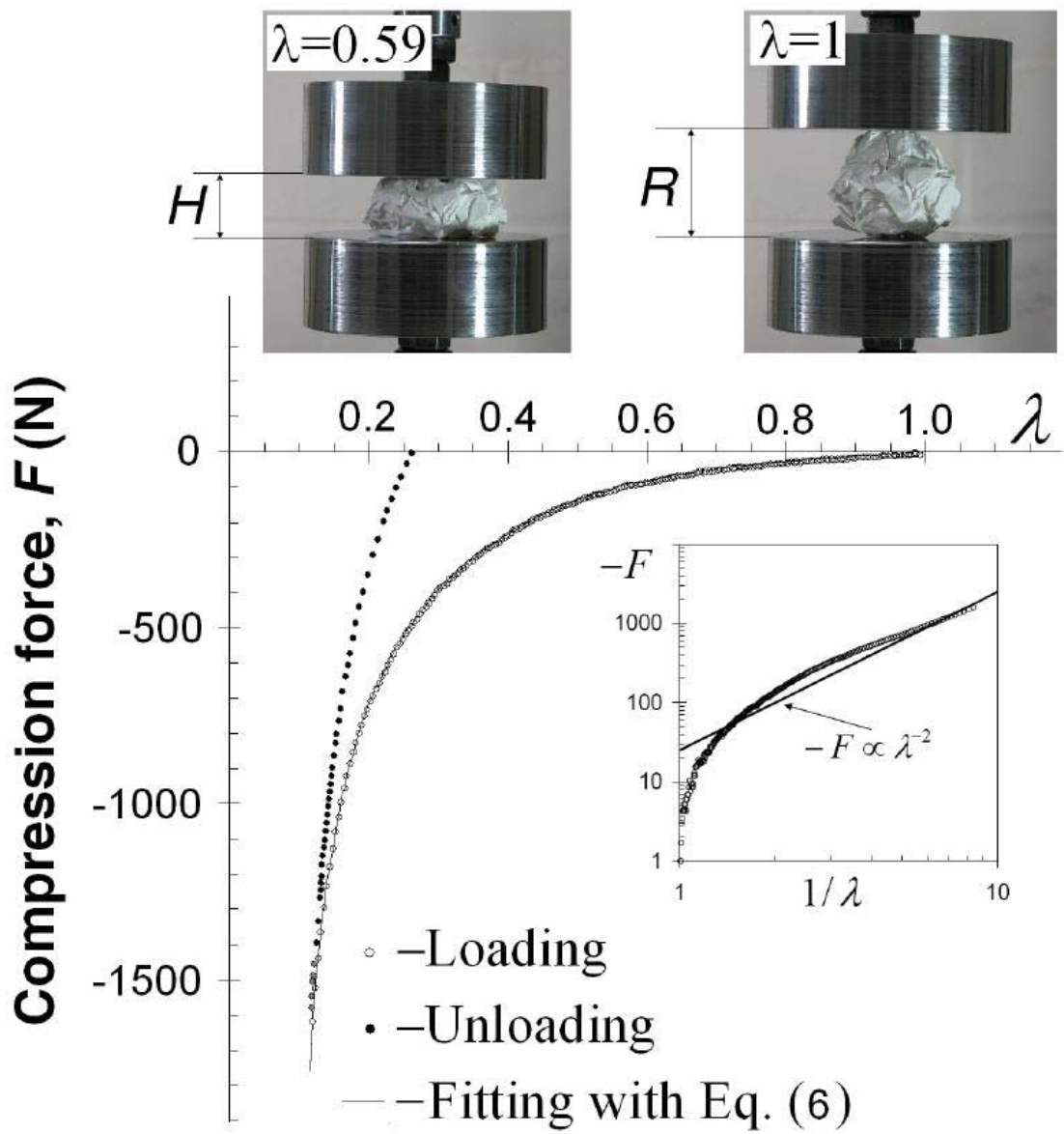

Figure 3.

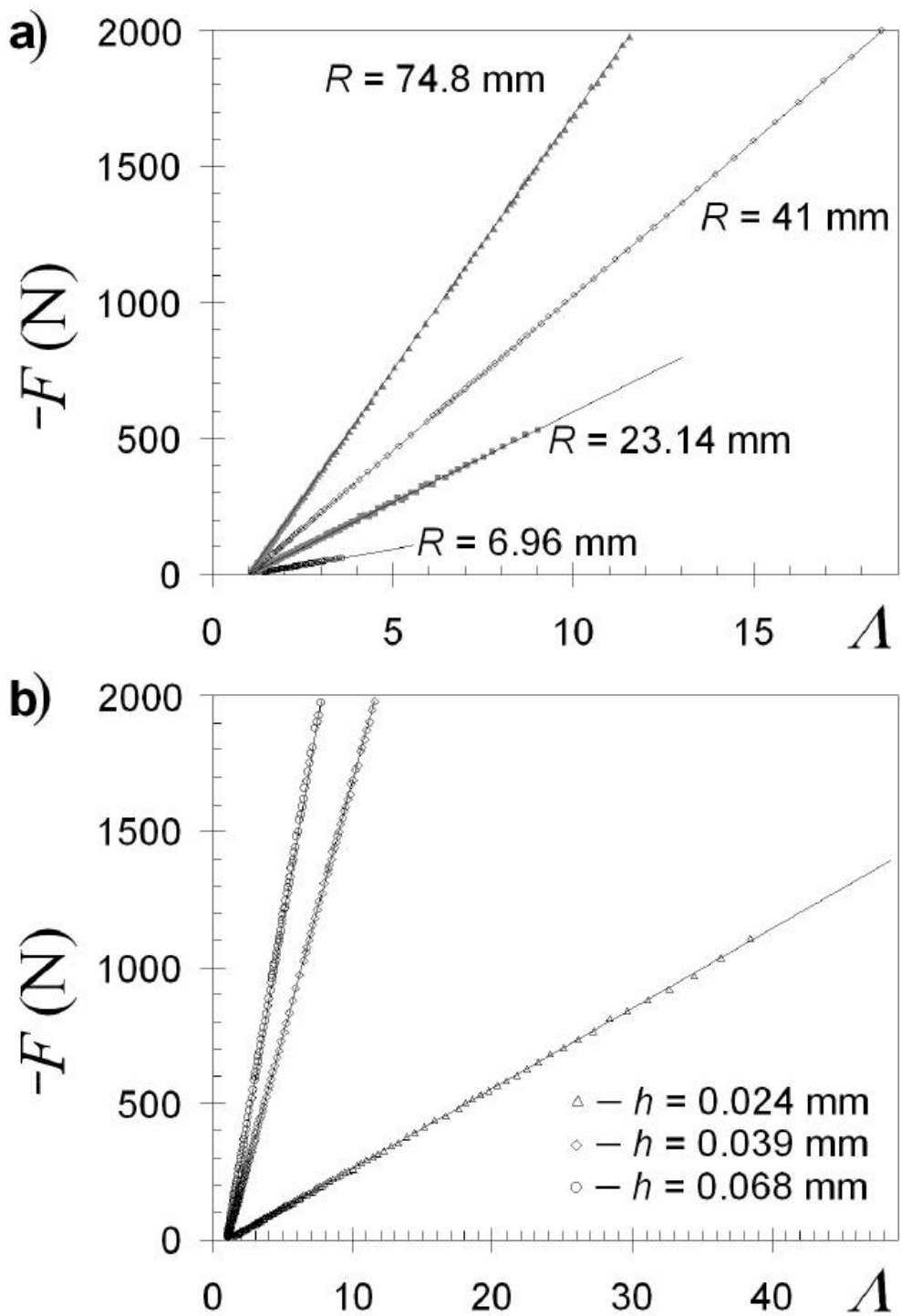

Figure 4.

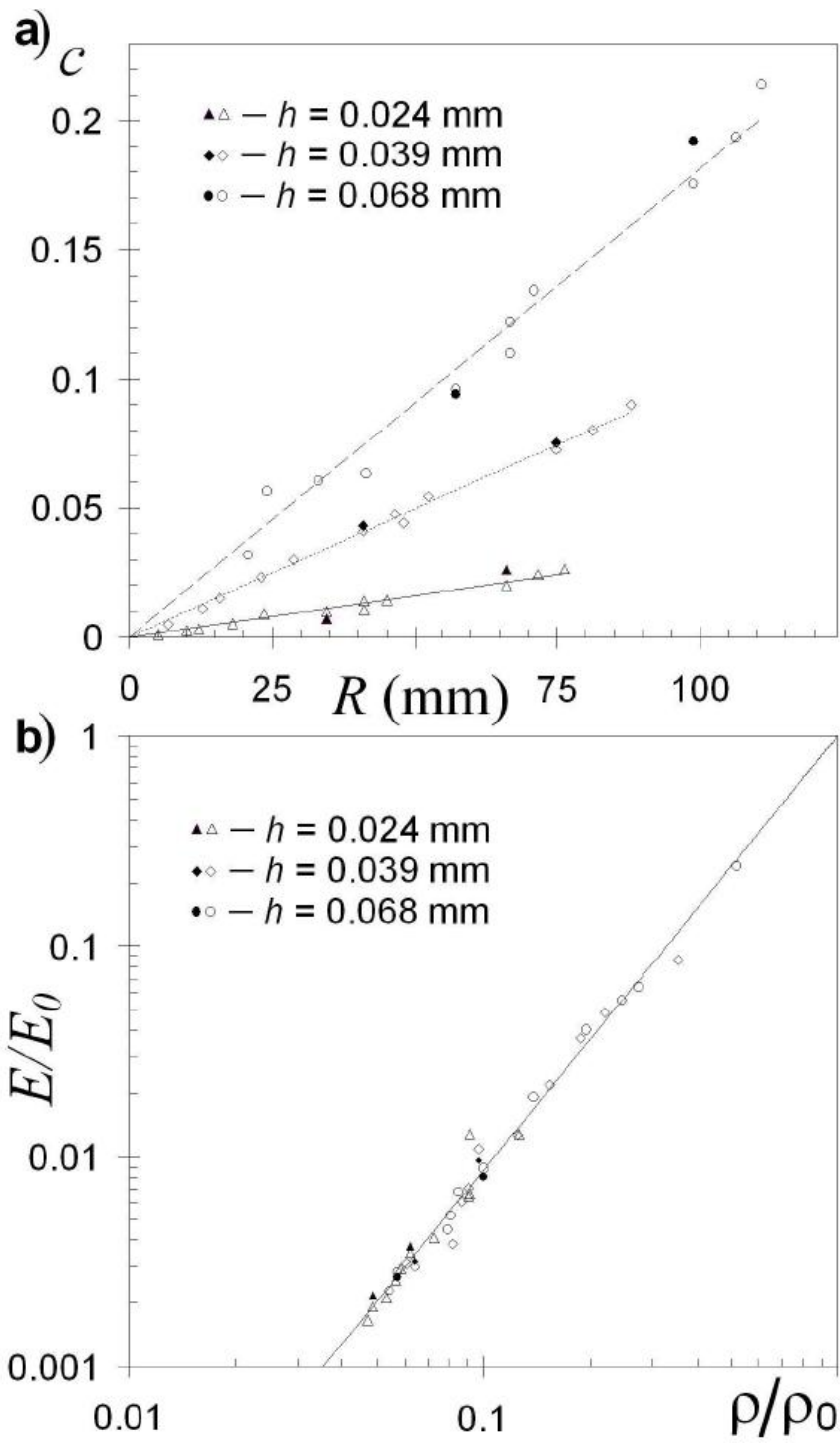

Figure 5.

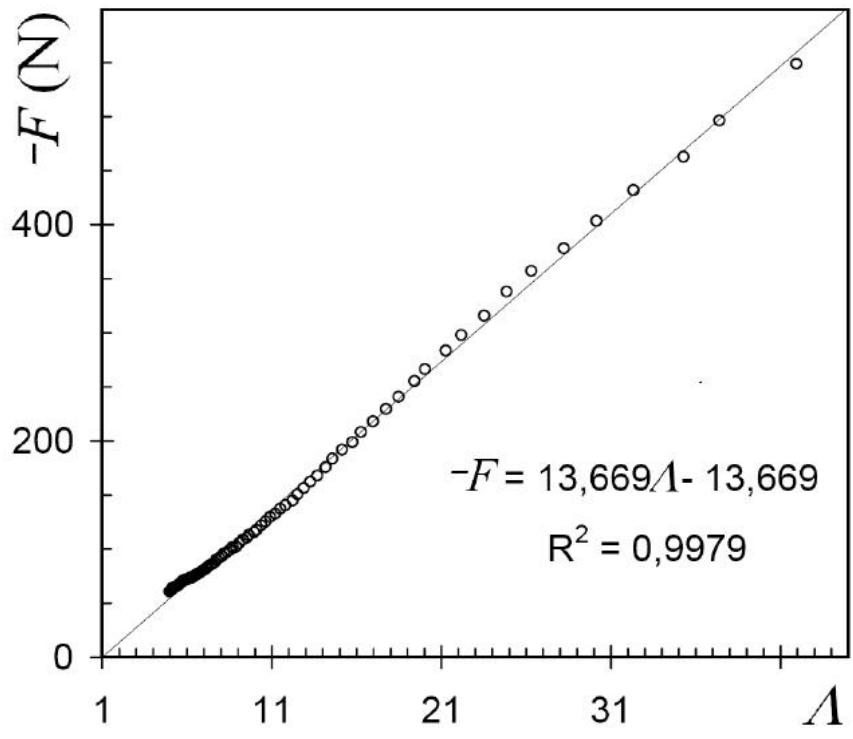

Figure 6.

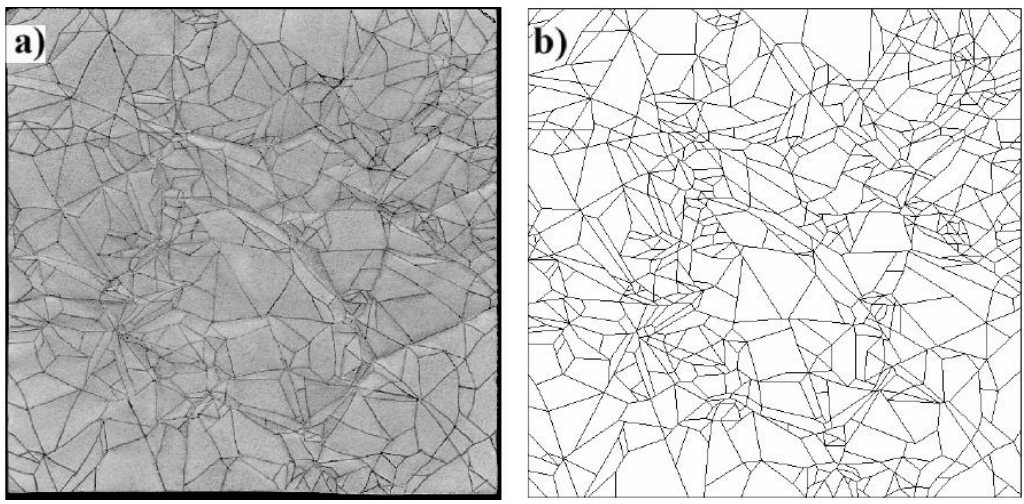

Figure 7.

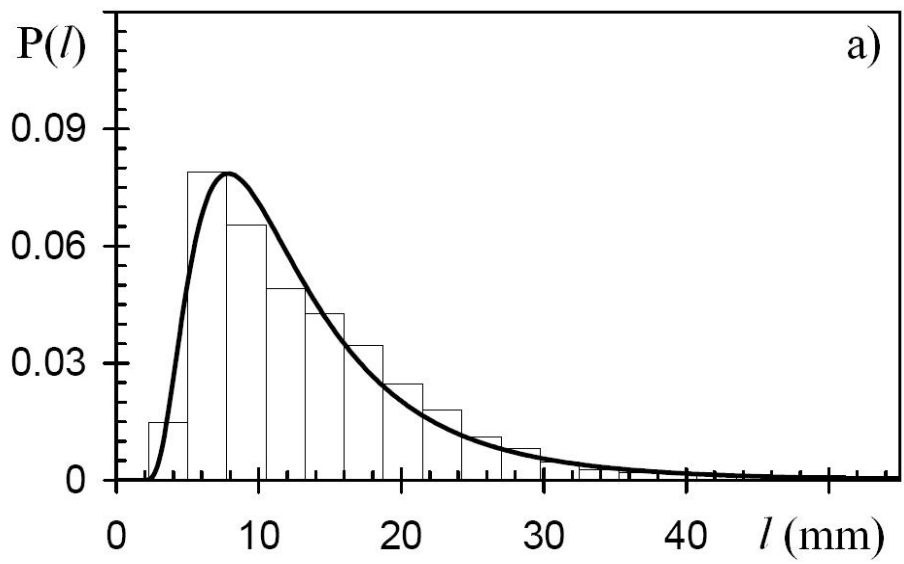

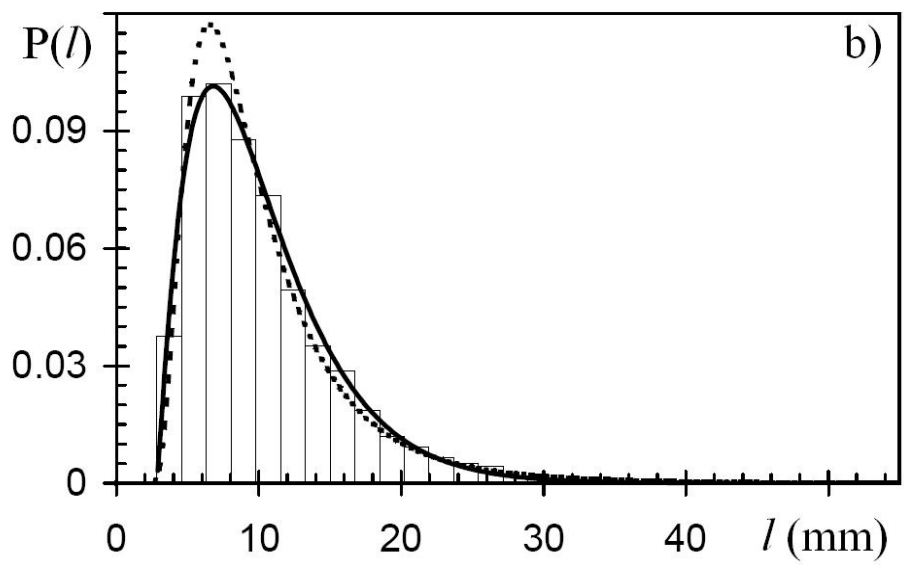

Figure 8.

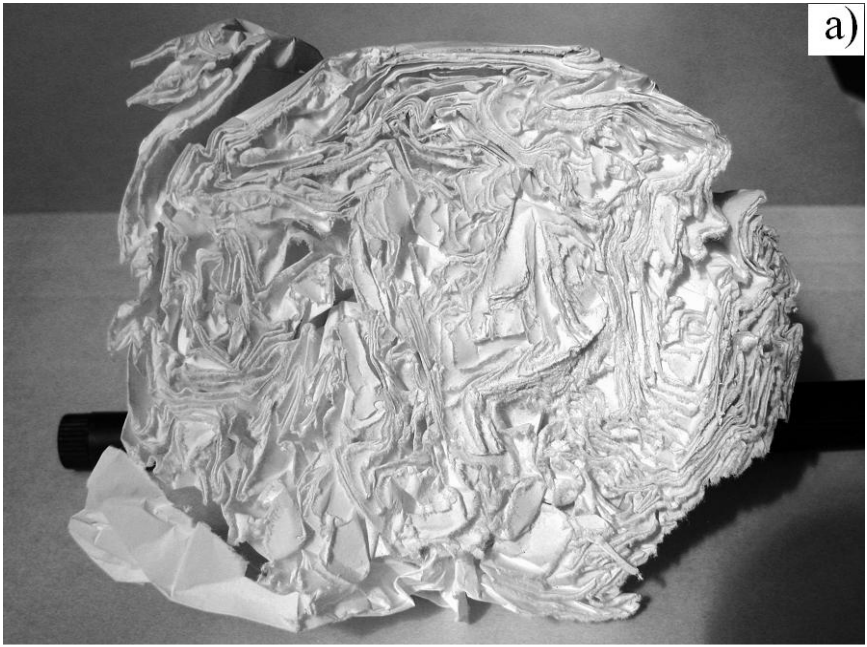

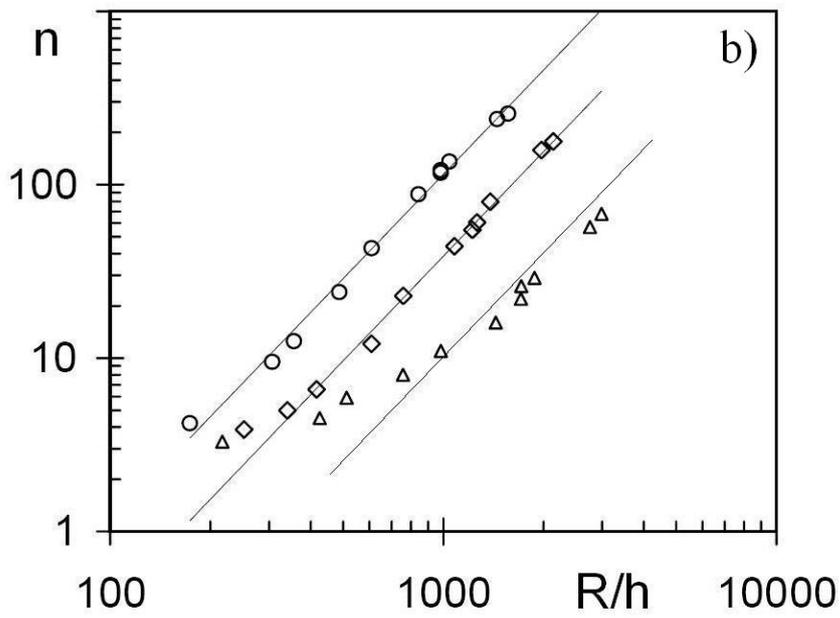

Figure 9.